\documentstyle[twoside,fleqn,espcrc2]{article}

\def\beq{\begin{equation}}
\def\eeq{\end{equation}}
\def\beqa{\begin{eqnarray}}
\def\eeqa{\end{eqnarray}}

\def\GeV{\nobreak\,\mbox{GeV}}
\def\fm{\nobreak\,\mbox{fm}}
\def\mb{\nobreak\,\mbox{mb}}

\newcommand{\AmS}{{\protect\the\textfont2
  A\kern-.1667em\lower.5ex\hbox{M}\kern-.125emS}}

\title{Pre-resonant Charmonium - Nucleon Cross Section 
       in the Model of the  Stochastic Vacuum}

\author{H.G. Dosch\address{Institut f\"ur Theoretische Physik, 
        Universit\"at Heidelberg
        \\
        Philosophenweg 16, D-6900 Heidelberg, Germany}, %
        	F.S. Navarra $^{\scriptsize {\mbox{b}}}$ and M. Nielsen 
\address{Instituto de F\'{\i}sica, Universidade de 
        S\~{a}o Paulo, \\
        C.P. 66318, 05315-970  S\~{a}o Paulo, SP, Brazil}}
\begin{document}

\begin{abstract}
We calculate the nonperturbative charmonium -  nucleon cross sections
with the model of the stochastic vacuum  which has been succesfully 
applied in many high energy reactions.  We also give a quantitative discussion 
of pre-resonance formation and medium effects. 
\end{abstract}

\maketitle

\section{Introduction}

Charmonium nucleon cross sections are of crucial importance in the 
context of Quark Gluon Plasma (QGP) physics \cite{QM97,QM99}.   
One needs to know the 
cross section $\sigma_{c  \overline c - N}$ in order 
to explain nuclear suppression  of $J/\Psi$ in terms of ordinary absorption 
by nucleons  without assuming a so called
``deconfining regime''. Estimates using perturbative QCD 
give values which are too
small to explain the observed absorption conventionally, but they are 
certainly not reliable for that genuine nonperturbative problem. A  
nonperturbative estimate may be
tried by applying vector dominance to 
$J/\Psi$  and  $\Psi'$ photoproduction.  
In this way a 
cross section of $\sigma_{J/\psi} \simeq 1.3$ mb for 
$\sqrt{s} \simeq 10$ GeV and 
$ \sigma_{\psi'}/\sigma_{J/\psi} \simeq 0.8$ has been obtained
\cite{HK,fs}. A more refined multichannel analysis \cite{HK} leads to  
$\sigma_{J/\psi} \simeq 3-4$ mb. 

The fact 
that the absorption cross section seems to be nearly the same both for 
$J/\psi$ and 
$\psi'$ has been interpreted as meaning that what is really absorbed is
rather a pre-resonant $c - \overline c$  state and not the physical 
particles. The size of this state has been estimated to be \cite{satz}
\beq
r_8 = \frac{1}{\sqrt{2 m_c \Lambda_{QCD}}} = 0.2 - 0.25 \mbox{fm}
\label{r8}
\eeq
and its cross section was then calculated with short distance QCD. A value of
$\sigma_8  \simeq 5.6 - 6.7$ mb was found.

In this note we calculate the pre-resonant $c - \overline c$ - nucleon 
cross sections in the  model of the stochastic vacuum (MSV)
\cite{Dosch:1987,Dosch:1988,Simonov:1988,dosch}. 
It has been applied to a large 
number of hadronic and photoproduction  processes with remarkably good 
success.

\section{The Model of the Stochastic Vacuum}
The basis of the MSV is the calculation of the scattering
amplitude of two colourless dipoles \cite{dgkp,dosch} based on a semiclassical
treatment developped by Nachtmann \cite{nach}. The
dipole-dipole scattering amplitude is  expressed as the expectation value of two
Wegner-Wilson loops with lightlike sides and  transversal extensions $\vec
r_{t 1}$ and $\vec r_{t 2}$ respectively. This leads to a  profile function $J(\vec b,
\vec r_{t 1},\vec r_{t 2})$ from which  hadron-hadron scattering amplitudes are
obtained by integrating over  different dipole sizes with the
transversal densities of the hadrons as weight functions according to 
\beqa
&& \sigma^{tot}_{(c - \overline c)-N} =  \int \, d^2 b \, d^2 r_{t 1} \,  d^2 r_{t 2}
\nonumber\\
&& \times  \,\,\,\,\,\,\,\, 
\rho_{(c - \overline c)-N}(\vec r_{t 1}) \,
\rho_N(\vec r_{t 2}) \,  J(\vec b, \vec r_{t 1},\vec r_{t 2})\; 
\label{totcross}
\eeqa 
Here $\rho_{(c - \overline c)- N}(\vec r_{t 1})$ and $\rho_N(\vec r_{t 2})$ are 
the transverse 
densities of the pre-resonant charmonium state and nucleon respectively. 

The basic ingredient of the model is the gauge invariant correlator of
two gluon field strength tensors. The latter is characterized by two
constants: the value at zero distance, the gluon condensate $<g^2FF>$,
and the correlation length $a$. We take these values from previous
applications of the model  \cite{dgkp} (and literature quoted there): 
$ <g^2FF>= 2.49 \,\,  \rm{ GeV}^4 $ and $ a= 0.346 \,\, \rm{ fm} $. 
The wave functions of the proton have been determined from
proton-proton and proton-antiproton scattering respectively. It turns out that 
the best description for the nucleon transverse density is given by that 
of a quark - diquark system with transversal distance ${\vec r}_t$ and density: 
\beq
\rho_N(\vec{r_t})=
|\Psi_p (\vec{r}_t)|^2= \frac{1}{2\pi}\frac{1}{S_p^2} \, e^{-\frac{
|\vec{r}_t|^2}{2S_p^2}}\; .
\label{Gausswfmeson}
\eeq
The value of the extension parameter,  $S_p=0.739\,\fm$, obtained from 
proton-proton scattering  agrees
very well with that obtained from the electromagnetic form factor in a similar 
treatment.

We start estimating the cross section in the case where the $c - \overline c$ pair 
is already in the physical $J/\psi$ or $\psi'$ states. The physical  wave functions
can be obtained in two different approaches:
\noindent
1) A numerical solution of the Schroedinger equation with the standard 
Cornell potential \cite{cornell}:
\beq
V = - \frac{4}{3} \frac{\alpha_s}{r} + \sigma r\;
\eeq

2) A Gaussian wave function determined by the electromagnetic decay
width of the $J/\Psi$ which has been used in a previous investigation of
$J/\Psi$ photoproduction \cite{dgkp}.

The linear potential can be calculated in the model
of the stochastic vacuum which  yields  the string tension:
\beq 
\sigma={8\kappa\over81\pi}<g^2 FF> a^2\; =0.179 \GeV^2\; 
\label{sigma}
\eeq
where the parameter $\kappa$  has been detemined in lattice calculations to be 
$\kappa = 0.8 $ \cite{lat}.

The other parameters, the charmed (constituent) mass and the
(frozen) strong coupling can be adjusted in order to give the correct
$J/\Psi$ and $\Psi'$ mass difference and the $J/\psi$ decay width:
\beq
m_c= 1.7 \rm{ GeV } \qquad  \alpha_s= 0.39 \; 
\label{malpha} 
\eeq
We also use the  standard Cornell model parameters \cite{cornell}: 
\beq
\alpha_s = 0.39 \,\, \sigma = 0.183 \, 
\GeV^2 \,\,  m_c = 1.84 \, \GeV \label{cornell} 
\eeq

From the numerical solution $\psi(|\vec r\,|)$ of the Schroedinger equation 
the transversal density is projected:
\beq
\rho_{J/\Psi} (\vec{r}_t) = \int\left|\psi(\sqrt{\vec{r}_t\,^2+ r_3^2})
\right|^2 dr_3\; 
\eeq
where $\vec{r}_t $ is the $J/\Psi$ transversal radius.

Given the values of $\alpha_s$,   $\sigma$  and  $m_c$  we solve 
the  non-relativistic Schroedinger equation numerically, obtain the wave 
function,  compute the transverse wave function and plugg it into the MSV 
calculation 
\cite{dosch}. 

In the pre-resonance absorption model, the pre-resonant 
charmonium state is either interpreted as a color-octet, $(c \overline c)_8$,  and 
a gluon in the hybrid  $(c \overline c)_8 - g$ state, or as a coherent  
$J/\Psi - \Psi'$ mixture. For the pre-resonant state 
we use a gaussian transverse wave function,  
as in Eq.~(\ref{Gausswfmeson}),  to represent a state 
with transversal radius $\sqrt{<r_t^2>} \, = \sqrt{2} S_{\psi}$. $ S_{\psi}$ 
is the  pre-resonance extension parameter analogous to $S_{p}$. The relation
between the average transverse radius and the average radius is given by: 
\beq
\sqrt{<r_t^2>} \simeq 0.82 \,\sqrt{<r^2>}  
\label{raios}
\eeq
With the knowledge of the wave functions and transformation properties of 
the constituents  we can compute the total cross 
section given by the MSV. The 
resulting nucleon - pre-resonant charmonium state cross section
 will be different if the pre-resonant charmonium state
consists of entities in the adjoint representation 
(as $(c \bar c)_8 - g$) or in the fundamental representation
(as a $J/\Psi - \Psi'$ mixture), the relation being:
\beq
\sigma_{\rm adjoint}
=\frac{2 N_C^2}{N_C^2-1}\sigma_{\rm fundamental}
\eeq
with $N_C=3$

\vskip 15mm

\section{Results}

The results are shown in Table I. In this table  
$\sqrt{<r^2>}$ is the root of the mean square distance of  quark and antiquark.
Wave function A) is the one obtained with the 
parameters given by  Eqs.~(\ref{sigma}) and (\ref{malpha}). Wave function B) 
corresponds to the standard Cornell model 
parameters,  Eq.~(\ref{cornell}). 
Wave function C) gives 
the result for the $J/\Psi-N$ cross section obtained with the 
weighted average of the longitudinally and transversely polarized 
$J/\Psi$ wave functions
from \cite{dgkp} with transversal sizes $\sqrt{<r_t^2>}=0.327\;\fm$ and
0.466 fm.

\vskip 5mm
\begin{center}
\begin{tabular}{||l|c|c|c||}
\hline
Wave function & $\sqrt{<r^2>}$ fm & $\sigma_{tot}$ [mb]\\
\hline\hline
$J/\Psi(1S)$&&\\
A &0.393&4.48\\
B  &0.375 &4.06\\
C &  & 4.69\\ \hline
$\Psi(2S)$&&\\
A:&0.788&17.9\\
\hline 
\end{tabular}
\end{center}
\begin{center}
\bf{TABLE I} {\small $J/\Psi-N$ and $\Psi'-N$ cross section} 
\end{center}
\vskip5mm

Averaging over our results for different wave functions,  
our final result for the $J/\Psi - N$ cross section is 
\beq
\sigma_{J/\psi - N}=4.4\pm0.6 \mb\;
\label{resu}
\eeq

The error is an estimate of uncertainties coming from the wave function and
the model. Other nonperturbative calculations of the
$J/\Psi - N$ cross section were presented in ref.\cite{gfssg98}, 
where the value  $\sigma_{J/\psi - N}=3.6$ mb was found, and in ref.\cite{huf}, 
which reported $\sigma_{J/\psi - N}=2.8$ mb. Our result is somewhat larger but 
still in agreement with these numbers. 
For $\Psi'$ our cross section is slightly smaller than 
$\sigma_{\psi'- N} =  20.0$ mb, obtained  in \cite{gfssg98} 
but larger than   $\sigma_{\psi'- N} =  10.5$ mb as found in 
\cite{huf}.

In Table II we show the results for the  absorption cross section of the 
pre-resonant charmonium state, interpreted as the  color-octet, 
$(c \overline c)_8 - g$  and as the coherent  
$J/\Psi - \Psi'$ mixture for  
different values of the average squared radius. 

\vspace{.5cm}
\begin{center}
\begin{tabular}{||c|c|c||}  \hline
$\sqrt{<r^2>}$ & $\sigma_{c \bar c}$ &$\sigma_{(c \bar c)_8-g}$  \\
(fm) & (mb) & (mb) \\
\hline\hline
0.24 & 1.79 &4.02 \\
\hline
0.31 & 2.76 &6.21\\
\hline
0.37 & 3.96 &8.91\\
\hline
0.43 & 5.30 &11.92\\
\hline
0.49 & 6.81 &15.32\\
\hline
0.55 & 8.50 &19.12\\
\hline
0.61 & 10.28 &23.13\\
\hline
\end{tabular}
\end{center}
\begin{center}
\bf{TABLE II} {\small The cross section pre-resonant charmonium-nucleon}
\end{center}
\vskip5mm

From our results we can see that a cross-section $\sigma_{\psi}^{abs} 
\simeq 6-7$ mb, needed to explain the $J/\Psi$ and $\Psi'$ suppression in
p-A collisions in the pre-resonance absorption model \cite{klns,spi}, 
is consistent with a pre-resonant charmonium state of 
size $\simeq 0.50-0.55$ fm if it is a $J/\Psi - \Psi'$ mixture 
or $\simeq 0.30-0.35$ fm for a $(c \bar c)_8 - g$ state. This last 
estimate can be compared with $r_8$ and $\sigma_8$ quoted above. For
$\sqrt{<r^2>} = r_8 \simeq 0.25$ fm  we obtain a cross section of $4$ mb 
instead of $6.7$ mb, as obtained in \cite{satz}. Inspite of the 
uncertainty in these numbers we can see that our calculation leads to 
smaller values for the cross sections. Alternatively, we may reverse the 
argument and say that the pre-resonant octet state must have a larger 
radius than previously estimated. This seems to be unlikely, especially 
in view of the estimates of sizes and lifetimes performed in \cite{huf}. 
This conclusion will become even 
stronger with the inclusion of medium effects.

In a hadronic medium, the QCD vacuum parameters  may change. Indeed,
lattice calculations \cite{lat} show that both
the correlation length and the quark and gluon condensates 
tend to decrease in a dense 
(or hot) medium. The first consequence, largely explored in cross section
calculations, is the change of hadron masses \cite{fe}. The second
consequence is a reduction of the string tension, $\sigma$, which 
leads to two competing effects,  which can be quantitatively compared in the
MSV. On one hand the cross section  tends to decrease strongly when the gluon
condensate or the correlation length  decrease. On the other hand, when the
string tension is reduced the  $c -  \overline c$ state becomes less confined
and will have a larger radius,  which, in turn, would lead to a larger cross
section for interactions with the  nucleons in the medium. In the MSV we
can determine which of these  effects is dominant.

Although all the calculations are done numerically,  we can parametrize the 
dependence of the cross sections on some specific quantities.  
We have therefore the following three possibilities to express the 
cross section as a function of the string tension, $\sigma$, the correlation 
length, $a$, and the gluon condensate, $<g^2FF>$ \cite{nos}:
\beq
\sigma_{\psi\,N} \propto \left\{\begin{array}{c}
\sigma^{5/6} a^{5/2}\\
 \sigma^{25/12} <g^2FF>^{-5/4}\\
 <g^2FF>^{5/6} a^{25/6}\end{array}\right. 
\; 
\label{fi}
\eeq

From the equation above we see that the final effect of the medium is a 
reduction in the cross section. Using the values  of the correlation length 
and the gluon condensate reduced
by 10\%: $a=0.31$ fm , $\langle g^2 FF\rangle=2.25$ GeV$^4$, we obtain a
40\% reduction in the cross sections. 

Taking this reduction into account 
the absorption cross sections obtained both for the physical $J/\psi$ 
(Eq. (\ref{resu})) and for the $J/\psi - \psi'$  mixture (second column 
in Table II) are smaller than the ones needed 
in Refs. \cite{huf} ,  \cite{klns} or \cite{spi} to explain experimental data.
The absorption cross section of the  hybrid  $(c \overline c)_8 - g$ state, 
even after the inclusion of medium effects, is still compatible (although 
somewhat small) with the values quoted in the mentioned papers.

To summarize, we calculated the nonperturbative $J/\Psi-N$ and $\Psi'-N$ 
cross sections with the MSV. We obtain
$\sigma_{J/\psi\, N}\sim4$ mb and $\sigma_{\psi'\,N}\sim 18$ mb. An interesting
prediction of the MSV is the strong depedence on the parameters of the QCD
vacuum which will most likely lead to a drastic reduction of them at higher
temperatures and perhaps also at higher densities.

\end{document}